\documentclass[a4paper,fleqn,usenatbib]{mnras}

\usepackage{newtxtext,newtxmath}

\usepackage[T1]{fontenc}
\usepackage{ae,aecompl}

\usepackage{graphicx}	
\usepackage{amsmath}	
\usepackage{amssymb}	

\title[SMC X-3: A giant X-ray outburst]{The 2016 super-Eddington outburst of SMC X-3: X-ray and optical properties and system parameters}

\author[L. J. Townsend et al.]{L. J. Townsend,$^{1}$\thanks{E-mail: townsend@ast.uct.ac.za (LJT)}
J. A. Kennea,$^{2}$
M. J. Coe,$^{3}$
V. A. McBride,$^{1,4}$
D. A. H. Buckley$^{4}$
\newauthor P. A. Evans,$^{5}$
A. Udalski$^{6}$
\\
$^{1}$Department of Astronomy, University of Cape Town, Private Bag X3, Rondebosch, 7701, South Africa\\
$^{2}$Department of Astronomy and Astrophysics, Pennsylvania State University, 525 Davey Lab, University Park, PA 16802, USA\\
$^{3}$Physics and Astronomy, University of Southampton, Southampton SO17 1BJ, UK\\
$^{4}$South African Astronomical Observatory, PO Box 9, Observatory 7935, South Africa\\
$^{5}$Department of Physics and Astronomy, University of Leicester, Leicester LE1 7RH, UK\\
$^{6}$Warsaw University Observatory, Aleje Ujazdowskie 4, PL-00-478 Warsaw, Poland
}

\date{Accepted 2017 July 20. Received 2017 July 20; in original form 2017 January 10}

\pubyear{2017}

\begin{document}
\label{firstpage}
\pagerange{\pageref{firstpage}--\pageref{lastpage}}
\maketitle

\begin{abstract}
On 2016 July 30 (MJD 57599), observations of the Small Magellanic Cloud by \textit{Swift}/XRT found an increase in X-ray counts coming from a position consistent with the Be/X-ray binary pulsar SMC X-3. Follow-up observations on 2016 August 3 (MJD 57603) and 2016 August 10 (MJD 57610) revealed a rapidly increasing count rate and confirmed the onset of a new X-ray outburst from the system. Further monitoring by \textit{Swift} began to uncover the enormity of the outburst, which peaked at $1.2\times10^{39}$\,erg/s on 2016 August 25 (MJD 57625). The system then began a gradual decline in flux that was still continuing over 5 months after the initial detection. We explore the X-ray and optical behaviour of SMC X-3 between 2016 July 30 and 2016 December 18 during this super-Eddington outburst. We apply a binary model to the spin-period evolution that takes into account the complex accretion changes over the outburst, to solve for the orbital parameters. Our results show SMC X-3 to be a system with a moderately low eccentricity amongst the Be/X-ray binary systems and to have a dynamically determined orbital period statistically consistent with the prominent period measured in the OGLE optical light curve. Our optical and X-ray derived ephemerides show that the peak in optical flux occurs roughly 6 days after periastron. The measured increase in I-band flux from the counterpart during the outburst is reflected in the measured equivalent width of the H$_{\alpha}$ line emission, though the H$_{\alpha}$ emission itself seems variable on sub-day time-scales, possibly due to the NS interacting with an inhomogeneous disc.
\end{abstract}




\section{Introduction}

High-mass X-ray binaries (HMXBs) comprise a compact object in orbit around an early-type star. Continuous or transient interactions between the two bodies cause X-rays to be produced on a range of time-scales and across a large range in luminosity. They have been observed in large numbers in the Milky Way and Magellanic Clouds, thanks to continuous monitoring from all-sky X-ray detectors and more focused surveys from several X-ray telescopes. The Small Magellanic Cloud (SMC) in particular is an exceptional host galaxy when it comes to HMXBs, with 121 high-confidence HMXB systems now known \citep{2016A&A...586A..81H}. This is higher than the known population of HMXBs in the Milky Way, despite the significantly lower mass of the SMC. The most probable reason for this is recent ($\sim$\,400\,Myr ago) bursts of star formation resulting in large numbers of massive, young stars, and the low line-of-sight extinction to the Magellanic Clouds. Work in this area is well documented (e.g. \citealt{2006MNRAS.370.2079D}; \citealt{2010ApJ...716L.140A}), though the details are beyond the scope of this paper. Of these 121 systems, 64 are confirmed pulsars (\citealt{2016A&A...586A..81H}; \citealt{2016ATel.9229....1V}). A detailed account of the optical and infra-red properties of this group of pulsating systems can be found in \cite{2015MNRAS.452..969C}.

HMXBs can be divided into 3 main groups: supergiant X-ray binaries (SGXRBs), Be-X/ray binaries (BeXRBs) and supergiant fast X-ray transients (SFXTs). As the focus of this paper is the well known BeXRB pulsar SMC X-3, we follow with a brief summary of this type of system. Recent good reviews of each type of HMXB system can be found in \cite{2011Ap&SS.332....1R} and \cite{2015A&ARv..23....2W}. BeXRBs comprise an early-type main-sequence or giant star that shows, or has shown, emission in the Balmer series. The majority of compact objects in these systems are neutron stars, though many remain unidentified. There is only one system with a confirmed black hole \citep{2014Natur.505..378C}. They are the most numerous group of HMXBs known and show significant variability in both time and luminosity, due to the wide range of observed binary parameters and the intrinsic variability of the Be star. In some circumstances, these systems can produce X-ray emission in excess of the Eddington limit for a nominal mass neutron star.

SMC X-3 was the third X-ray source to be discovered in the SMC, by the SAS 3 satellite in 1977 (\citealt{1977IAUC.3125....1L}; \citealt{1978ApJ...221L..37C}). In 2002, the \textit{Rossi X-ray Timing Explorer (RXTE)} discovered a new pulsating X-ray source in the SMC, with a period of roughly 7.8\,s \citep{2003HEAD....7.1730C}. Those authors detected this period several times during the following year, leading to the identification of a probable orbital period of $\sim$45\,d. It was not until 2004 that the association between this X-ray pulsar and SMC X-3 was made. \cite{2004ATel..225....1E} used an archival \textit{Chandra} observation to precisely locate SMC X-3 and determine a 7.78\,s period in the light curve. The observation was made three days before a \textit{RXTE} detection of the 7.8\,s period, confirming the association of the \textit{RXTE} and \textit{Chandra} sources. The precise \textit{Chandra} position from this work (00h52m05.7s, -72d26m05s (J2000.0), uncertainty 0.6\,arcsec) was consistent with the position of the originally proposed optical counterpart to SMC X-3 (\citealt{1977IAUC.3134....3V}; \citealt{1978ApJ...223L..79C}). An optical modulation in long-term MACHO photometry was found at 44.86\,d by \cite{2004AJ....128..709C}. The authors note that this is very close to the X-ray period determined by \cite{2003HEAD....7.1730C} and conclude that it represents the binary period of the system. \cite{2005PhDT.........3E} also found a significant modulation in long-term OGLE photometry at 44.8\,d, during a phase of apparent X-ray quiescence. The long-term \textit{RXTE} light curve of SMC X-3 is presented in \cite{2008ApJS..177..189G}. The authors determine a full X-ray ephemeris of MJD 52,250.9 $\pm$ 1.4 $+$ 44.92n $\pm$ 0.06\,d, where 44.92\,d is the prominent modulation in the light curve and agrees well with the optically derived periods. In addition, the light curve seems to show distinct periods of X-ray activity lasting between $\sim$\,200--400\,d and separated by $\sim$\,1000\,d. Intriguingly, the authors note that there appears to be evidence of weak X-ray emission at apastron in the folded light curve and that the overall trend in the light curve is a spin-down of the pulsar. The \textit{RXTE} light curve and spin period evolution are discussed in more detail in section 2.3. \cite{2013MNRAS.431..252S} refine the optical period to 44.94$\pm$0.02\,d using more recent OGLE III data. The authors also observe a dip in the folded light curve immediately before optical maximum, when the system is in an optically bright state. Finally, the spectral type of the counterpart was determined to be B1--1.5\,IV--V \citep{2008MNRAS.388.1198M}.

Here we describe the recent super-Eddington outburst of SMC X-3. The original detection of the outburst was made by \textit{MAXI} on 2016 August 08 \citep{2016ATel.9348....1N}, although they were unable to identify conclusively that it was SMC X-3. The association was confirmed from multiple detections of the outburst by \textit{Swift} \citep{2016ATel.9362....1K}. The outburst was estimated to begin sometime between 2016 July 16 and 2016 July 30 based on \textit{Swift/XRT} observations of the SMC, and was still ongoing over 5 months after this initial detection. After submission of this manuscript, further analysis of this outburst was presented in \cite{2017arXiv170102983W} and \cite{2017arXiv170200966T}. Those authors independently derive binary solutions for SMC X-3, which we compare to our results. They also present determinations of the magnetic field strength of the neutron star, which are slightly inconsistent with each other. The magnetic field was previously estimated by \cite{2014MNRAS.437.3863K}, based on the long-term spin period evolution of the neutron star. Those results are broadly consistent with the value found in \cite{2017arXiv170200966T}.

In section 2, we present the \textit{Swift} X-ray observations of the outburst between 2016 July 30 and 2016 December 18 and historical \textit{RXTE} measurements of the long-term spin period. In section 3, we present the optical photometric and spectroscopic observations. The model used to determine the binary solution and mass accretion rate is described in section 4 and in section 5 we discuss the significance of this outburst and draw together our conclusions based on the combined optical and X-ray activity.

\section{X-ray observations}

In this section, we present the \textit{Swift} BAT and XRT light curves and XRT timing of the 2016 super-Eddington outburst of SMC X-3. The \textit{Swift} timing information is compared to the \textit{RXTE} long-term spin period history.

\begin{figure}
	\vspace{-0.8cm}
	\hspace{-0.5cm}
	\includegraphics[width=1.17\columnwidth,angle=0]{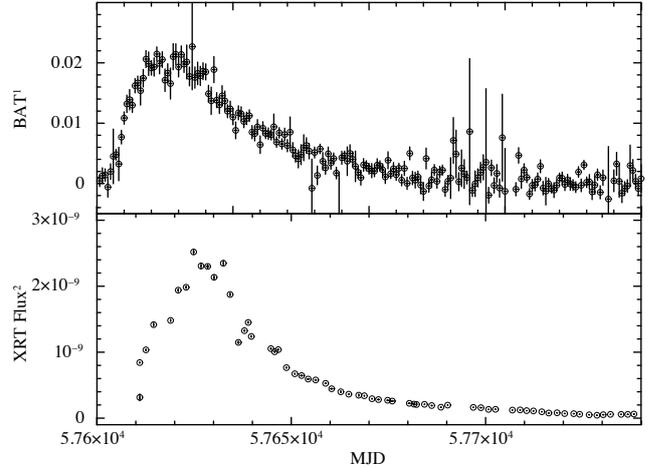}
	\vspace{-0.5cm}
    \caption{\textit{Top panel}: \textit{Swift}/BAT light curve of the 2016 outburst of SMC X-3. The BAT count rate is in units of counts/s/detector as reported by the BAT Transient Monitor \citep{2013ApJS..209...14K}. \textit{Bottom panel}: \textit{Swift}/XRT light curve of the 2016 outburst of SMC X-3. The XRT flux is in units of erg/s/cm$^{2}$ in the 0.5--10\,keV range and has been calculated from spectral fitting of each observation. The error bars are 1\,$\sigma$ in both panels.
}
    \label{fig:swiftlc}
\end{figure}

\begin{figure}
	\vspace{-0.8cm}
	\hspace{-0.5cm}
	\includegraphics[width=1.17\columnwidth,angle=0]{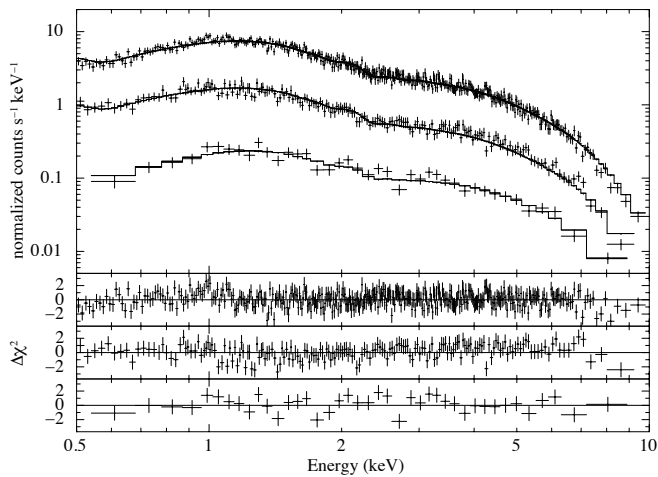}
	\vspace{-0.5cm}
        \caption{Swift/XRT spectral fits at different flux levels from three epochs of the SMC X-3 outburst. These data are fit with a simple absorbed power-law model, with a low absorption column ($N_\mathrm{H} \simeq 6.1\times10^{20}~\mathrm{cm^{-2}}$). In order of decreasing flux, the dates of these observations are 2016 August 20, 2016 October 9 and 2016 December 12. These spectra are binned to have a minimum of 50 counts per bin, the power-law model fits in all cases with a reduced $\chi^2 \simeq 1$. No additional model components are  required to describe the XRT data at any point during the period of outburst covered in this paper, there is no evidence of any thermal component.
}
    \label{fig:swiftspec}
\end{figure}

\subsection{Swift}


The \textit{Swift}/XRT began observing the X-ray outburst of SMC X-3 on 2016 July 30 and continued to detect emission over 5 months later. We present data up to 2016 December 18, which are summarised in Appendix A. Fig.~\ref{fig:swiftlc} shows the combined BAT and XRT light curves during the outburst. The BAT light curve shown in the top panel is as reported by the BAT Transient Monitor \citep{2013ApJS..209...14K}. The XRT fluxes shown in the bottom panel were found using time resolved spectral fitting to individual XRT observations listed in Appendix A.

XRT data were analysed utilizing the HEAsoft v6.19, and the latest XRT calibration files (CALDB 20160609). All XRT data were reprocessed with the standard \textit{xrtpipeline} software, with exposure maps generated in order to compensate for the presence of hot-columns. In order to avoid issues of pile-up, and to use data which has sufficient timing resolution to detect pulsations, we use only data taken in Windowed Timing (WT) mode, in the energy range of $0.5 - 10$~keV. In addition, to avoid issues of event redistribution, we only utilize WT events of grade 0 for our spectral analysis. Events were extracted from a region of radius 20 pixels around the optical position of SMC X-3, with a background taken from an annulus between 90 and 110 pixels away from the central region, in order to ensure the smallest contamination from SMC X-3 itself.

After careful handing of issues of calibration, pile-up and background subtraction, we found that the XRT spectra throughout the reported outburst period can be well described by a single power-law model, with more complex models not being statistically required. Although no other components are required to fit the spectrum, SMC X-3 does show  spectral variability over the outburst, with the photon index averaging $\Gamma\sim1.08$ during the brightest part of the outburst, and hardening to an average of $\Gamma\sim~0.85$ after MJD 57700. Fig.~\ref{fig:swiftspec} shows spectral fits of XRT WT data at epochs representing three different flux levels, all well fit by a simple power-law model.
This power-law model was used to calculate both the fluxes shown in Fig.\ref{fig:swiftlc} and luminosities used in modelling the expected accretion driven spin-up in section 4. Fluxes were obtained utilizing the XSPEC \texttt{cflux} model component for the range of $0.5 - 10$~keV.

We note that analysis of WT data can be complex due to issues with correctly subtracting the background, in light of the compresssed 1-dimensional nature of the data. Incorrect subtraction of the background can often lead to the appearance of a false soft component in the data, which would become more dominant as SMC X-3 fades. The presence of hot columns that are masked out in XRT data further complicates background subtraction, as failing to compensate for this can lead to an over-subtraction of background. In addition, the issue of event redistribution can cause a false soft component, if the incorrect or out-of-date calibration files are used\footnote{http://www.swift.ac.uk/analysis/xrt/digest\_cal.php\#abs}. As event redistribution tends to occur mainly in higher grade events, this issue can be mostly avoided by not using WT grade 1 and 2 events. False soft components have often been mistaken as a thermal component in low-mass X-ray binary (LMXB) systems. We suspect that one of these issues is the source of the thermal excess reported in \cite{2017arXiv170102983W}, as we find no evidence of it in our analysis. Furthermore, we suspect the reported soft component seen in EPIC-pn timing mode data on SMC X-3 \citep{2017arXiv170102983W} may be incorrect due to known issues of calibration of that mode at low energies (e.g. Pintore et al., 2014).

The power law index and absorption are typical for X-ray binaries in the SMC and do not show anything abnormal during this giant outburst. Using the flux from the spectral fits, we estimate the outburst reached a peak luminosity of $1.16\times10^{39}$\,erg/s (0.5 -- 10 keV), assuming a source distance of 62\,kpc\footnote{\cite{2016ApJ...816...49S} showed that the line-of-sight depth across the SMC varies around the average from -10\,kpc in the East, to +10\,kpc in the West. SMC X-3 lies roughly in the centre of this distribution and is therefore less likely to be at a significantly different distance than the average value.} (\citealt{2016ApJ...816...49S} and references therein). This is above the Eddington limit for a 1.4\,M$_{\odot}$ neutron star by a factor of roughly 6. As can be seen in Fig.~\ref{fig:swiftlc}, SMC X-3 underwent a rapid increase in flux, before plateauing briefly at this super-Eddington luminosity and entering a phase of more gradual decline.

For the timing analysis, we used event data which were barycentrically corrected using the standard \texttt{barycorr} command. Period searches were performed utilizing time tagged events from WT data, which have a time resolution of 1.7791\,ms. Each observation made with the XRT in WT mode was run through a timing analysis pipeline to search for the spin period of the neutron star. The period was determined using a standard Rayleighs $Z_{N}^{2}$ search \citep{1983A&A...128..245B}, searching a period space around the known 7.78\,s period of the source at a resolution of 10$^{-6}$\,s. As the pulse profile for SMC X-3 is double peaked, and close to sinusoidal, we found that performing a $Z^2_N$ search with $N=2$ was sufficient.

Error estimation was performed using the Monte-Carlo method of \cite{1999ApJ...522L..49G}, where the folded pulse profile is fitted at the determined frequency and then used to generate 500 simulated event files with Poissonian statistics, utilizing the same observing windows and exposures as the real observations, on which an identical period search is performed again. The error is then calculated by fitting a Gaussian model to the distribution of the fitted periods found from these faked datasets.

The accuracy of \textit{Swift}/XRT timing utilising these data has been accurately verified by \citep{2012A&A...548A..28C}. The Monte-Carlo simulations generate event files that exactly mimic the format, time resolution and sampling rate of WT events in real data, with an assumption that the pulse profile and the DC level of the source does not vary significantly across the observation. Given that the observations are typically short, we believe that this is a reasonable approximation. However, if the approximation is not true, it is likely that the effect of this would be to underestimate the error on the period. In all simulations, the simulated histogram of periods was found to be distributed in a Gaussian fashion. All period searching (on real and simulated data) is done with a fixed period resolution (1 x 10$^{-6}$\,s), which was chosen as it is both adequate to sample for Doppler shifts and because the error on each measurement is typically at least an order of magnitude larger. There is no evidence of any red-noise effects due to how we sample the data, real or simulated.

\subsection{RXTE}

\begin{figure}
	\includegraphics[width=1.05\columnwidth,angle=0]{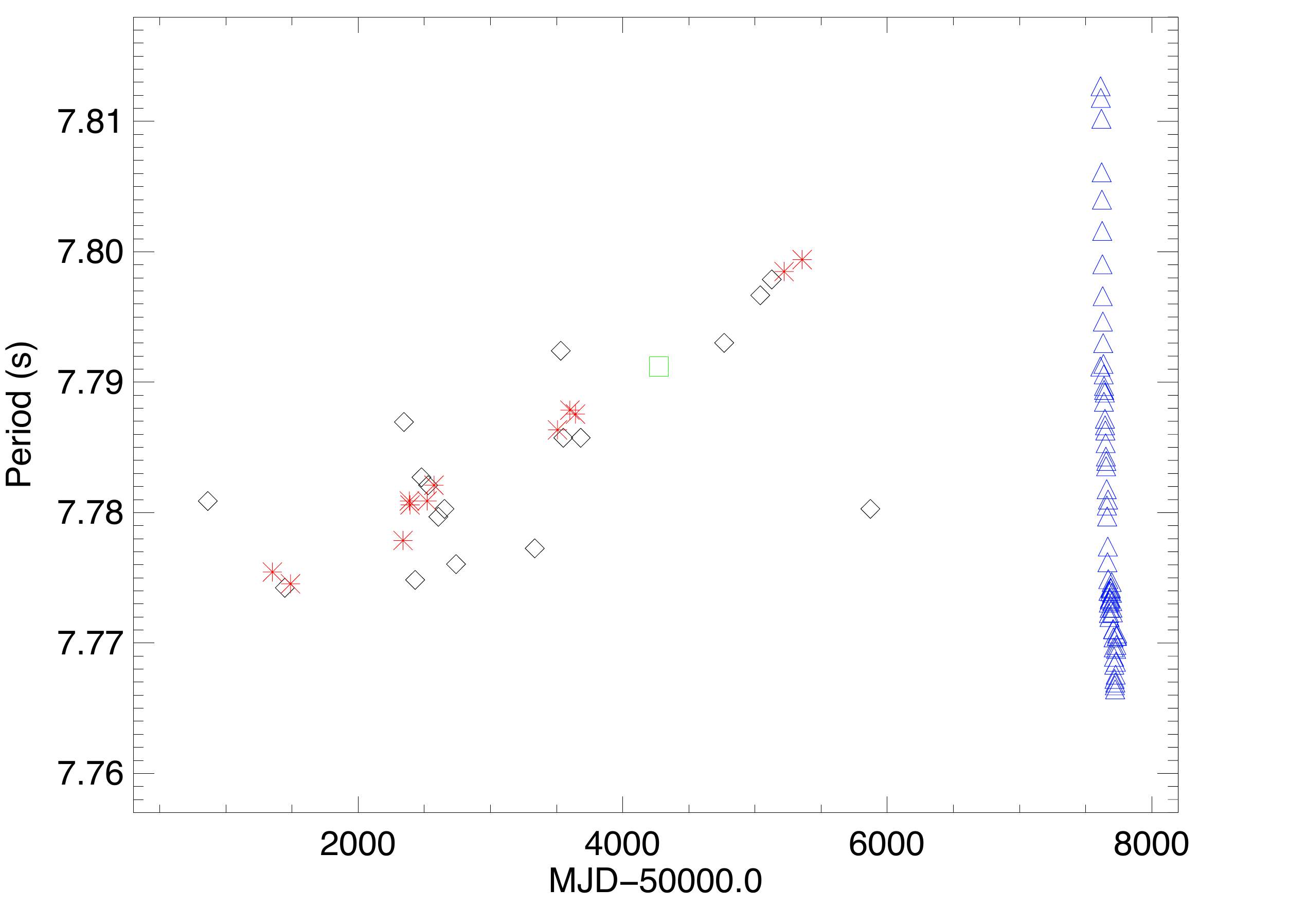}
    \caption{X-ray derived pulsed period history of SMC X-3. Black diamonds and red stars denote \textit{RXTE} period detections above the 99\% and 99.99\% confidence levels respectively. Blue triangles denote \textit{Swift} detections of the pulse period during the current outburst. A single \textit{XMM-Newton} detection at MJD 54274 was found in the literature and is denoted by a green square.}
    \label{fig:rxte}
\end{figure}

The position of SMC X-3 was observed by \textit{RXTE} on a roughly weekly basis between 1999 and the end of the mission in 2012. As described in section 1, the eventual detection of a 7.78\,s period by \textit{RXTE} in 2002 led to the discovery that the known X-ray source, SMC X-3, was an accreting neutron star. In Fig.~\ref{fig:rxte}, we show the full spin period history of SMC X-3 as seen by \textit{RXTE}. This figure is an extension of the spin period panel of Fig. 7 in \cite{2008ApJS..177..189G}. The period measurements were calculated from individual exposures of the \textit{RXTE}/PCA using a Lomb-Scargle analysis method. Because of the large field of view and non-imaging of the PCA, we use the method described in \cite{2008ApJS..177..189G} to subtract out each detected periodicity from the periodogram before searching for the next. The periodogram is searched for known and unknown periods by performing a global period search and smaller searches in a range around a known pulsar period. The detections of SMC X-3 shown in Fig.~\ref{fig:rxte} were found using this search method. More details of this analysis can be found in \cite{2008ApJS..177..189G}. In addition, we have included the period detections made by \textit{Swift} during the current giant outburst and a single \textit{XMM-Newton} period measurement found in the literature \citep{2008A&A...489..327H}. One can see the obvious difference between the historical spin period evolution and the evolution during the giant outburst. This is discussed further in section 5.

\section{Optical observations}

In this section, we present optical photometry from the Optical Gravitational Lensing Experiment (OGLE) and optical spectroscopy from the Southern African Large Telescope (SALT) and South African Astronomical Observatory (SAAO) 1.9\,m telescope in South Africa, and the European Southern Observatory (ESO) 3.6\,m telescope and New Technology Telescope (NTT) in Chile. We use the data to provide the most precise optical period measurement to date and show the evolution of the circumstellar disc over a long baseline, linking this to the X-ray activity in this system.

\begin{figure*}
	\includegraphics[width=1.5\columnwidth,angle=90]{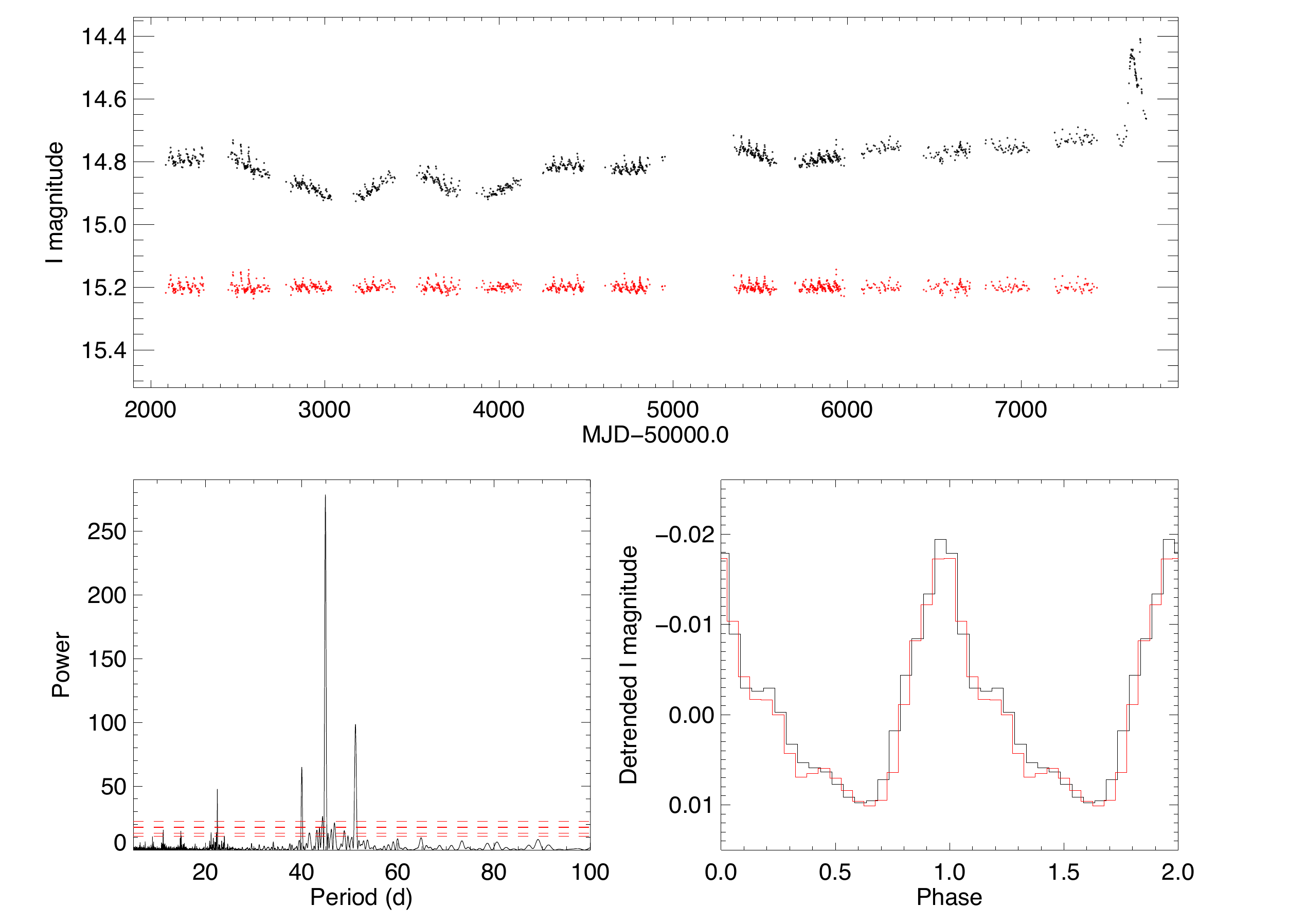}
    \caption{\textit{Top panel}: OGLE III \& IV light curve of SMC X-3 in the I-band (black). The detrended light curve is shown for comparison (red). For clarity, 15.2 magnitudes have been added to the detrended light curve. \textit{Bottom-left panel}: Lomb-Scargle periodogram of the detrended OGLE light curve, excluding the recent outburst. The most prominent peak corresponds to a period of 44.918\,d. The next highest peaks at periods of 40.00\,d and 51.16\,d are attributed to aliasing. There are smaller peaks visible, corresponding to harmonics of the main period. The dashed red lines are the 90, 99, 99.99 and 99.9999\% confidence levels. \textit{Bottom-right panel}: The detrended OGLE light curve folded at the 44.918\,d period. OGLE III is in black and OGLE IV in red.}
    \label{fig:ogle}
\end{figure*}

\subsection{OGLE photometry}

SMC X-3 was observed during the third and fourth phases of the OGLE project, which has been taking V and I band images with the 1.3 m Warsaw telescope at the Las Campanas Observatory, Chile for the past 20 years (\citealt{2008AcA....58...69U}; \citealt{2015AcA....65....1U}). The cadence of observations vary from a few days to roughly a week in the I band and slightly longer in the V band. The I band is a much better tracer of the circumstellar disc in BeXRB systems and, being of high cadence, it is ideal to show the optical variability in SMC X-3. The full 14 year OGLE light curve of SMC X-3 is shown in the top panel of Fig.~\ref{fig:ogle}. The black curve shows the combined OGLE III \& IV data, whilst the red curve shows this light curve detrended with a 101\,d sliding window function to remove the longer-term aperiodic variability, likely caused by changes in the size of the Be star disc. We manually removed the large outburst at the end of the light curve before detrending. The periodic modulations that can be seen throughout both light curves as regular spikes are typically thought to represent the orbital period of the system, though it is becoming clearer that this may not be entirely accurate (see discussion section). A large 0.4 magnitude jump in flux near the end of the light curve coincides with the current giant X-ray outburst. It is likely that a sudden growth in the disc has caused this and triggered extensive accretion onto the orbiting neutron star. The optical outburst began on or just after MJD 57601, approximately 2 days after the first X-ray detection by \textit{Swift}, though it is difficult to determine the precise onset due to the cadence of the light curve as well as the aforementioned uncertainty in the onset of the X-ray outburst.

The lower-left panel of Fig.~\ref{fig:ogle} shows the Lomb-Scargle periodogram of the detrended light curve. Over-plotted in dashed red lines are the 90, 99, 99.99 and 99.9999\% confidence levels, computed directly from the power spectrum statistics. Most of the power is seen at a period of 44.918 $\pm$ 0.009\,d., which agrees well with the optical value determined by \cite{2013MNRAS.431..252S} and the X-ray value determined by \cite{2008ApJS..177..189G}. The uncertainty is calculated based on the formula for the standard deviation of the frequency given in \cite{1986ApJ...302..757H}. The first harmonic of the fundamental period is easily seen at 22.46\,d. Higher order harmonics are also detected down to the sixth harmonic around 6.40\,d. This results in quite a complex orbital profile (lower-right panel of Fig.~\ref{fig:ogle}). The OGLE III and IV light curves were folded separately and plotted for comparison. One can see that there is very little difference between the two phases, indicating that the period is very stable over more than a decade of observations. The profiles are also highly skewed in a 'fast rise, exponential decay' shape. This shape is indicative of the modulation being related to the binary period of the system (e.g. \citealt{2012MNRAS.423.3663B}). The ephemeris of peak flux is MJD 57682.14\,$\pm$\,0.37. This will be discussed further in section 5 in the context of the orbital ephemeris from binary model fitting.

\subsection{Spectroscopy}

Observations were obtained from the SALT and SAAO 1.9\,m telescopes just after the peak of the X-ray outburst to measure the shape and equivalent width of the H$_{\alpha}$ emission line, and to see if any other features were present. These were compared with several archival spectra from the SAAO 1.9\,m and ESO 3.6\,m and NTT telescopes obtained sporadically over the past $\sim$20 years. The details of these observations, and the measured equivalent widths, are shown in Table \ref{tab:spectra}.

\begin{table}
	\centering
	\caption{Observation time, telescope and equivalent width of each spectrum used in our analysis.}
	\label{tab:spectra}
	\begin{tabular}{|l|l|l|l|}
  		\hline
  		Date & MJD & Telescope & H$_{\alpha}$ EW \\
  		\hline
  		\hline
  		11 Nov 2001 & 52224 & SAAO 1.9m & -11.4 $\pm$ 0.5 \\
  		06 Nov 2003 & 52949 & SAAO 1.9m & -12.1 $\pm$ 0.7 \\
		27 Oct 2005 & 53670 & SAAO 1.9m & -11.8 $\pm$ 1.0 \\
  		10 Nov 2006 & 54049 & SAAO 1.9m & -10.2 $\pm$ 0.4 \\
  		17 Sept 2007 & 54360 & ESO 3.6m & -11.7 $\pm$ 0.3 \\
  		11 Dec 2011 & 55906 & NTT & -12.4 $\pm$ 0.2 \\

  		04 Sept 2016 & 57635 & SAAO 1.9m & -13.5 $\pm$ 0.7 \\
  		06 Sept 2016 & 57637 & SAAO 1.9m & -12.7 $\pm$ 0.6 \\
  		06 Sept 2016 & 57637 & SALT & -11.5 $\pm$ 0.1 \\
  		11 Sept 2016 & 57642 & SALT & -11.0 $\pm$ 0.1 \\
  		13 Nov 2016 & 57705 & SALT & -13.4 $\pm$ 0.1 \\
  		25 Nov 2016 & 57717 & SAAO 1.9m & -15.5 $\pm$ 0.4 \\
 		\hline
	\end{tabular}
\end{table}

\begin{figure}
	\vspace{-1.4cm}
	\hspace{-0.7cm}
	\includegraphics[width=1.1\columnwidth,angle=0]{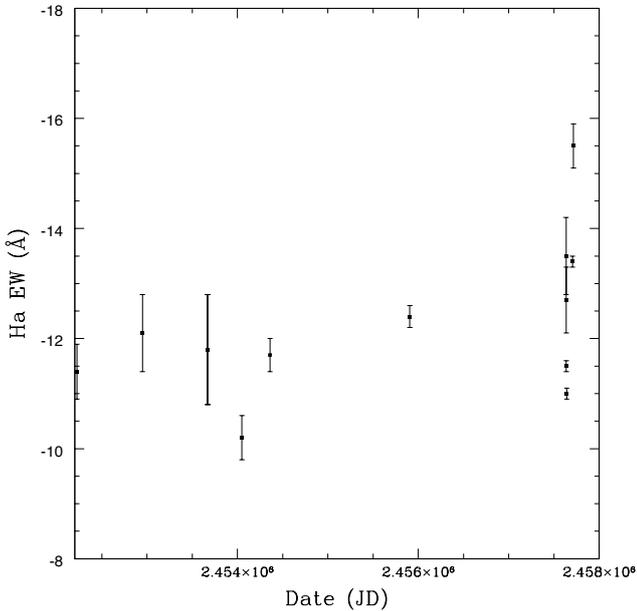}
	\vspace{-2.5cm}
    \caption{Measured values of the H$_{\alpha}$ equivalent width from the spectra listed in Table \ref{tab:spectra}.}
    \label{fig:halpha}
\end{figure}

Fig.~\ref{fig:halpha} shows the H$_{\alpha}$ equivalent width measurement from each of the archival spectra obtained with the SAAO 1.9\,m telescope and the ESO 3.6\,m and NTT telescopes, as well as the SAAO 1.9\,m and SALT/RSS spectra taken during the current outburst. Despite large gaps in the coverage of SMC X-3, there is moderate evidence that the circumstellar disc around the Be star has grown in recent months; as suggested by the OGLE photometry. However, there is also evidence for rapid variability in the line emission that is unexpected in this type of binary system, even during a large outburst. This point is discussed in section 5.

\begin{figure}
		\vspace{-1.4cm}
		\hspace{-1.0cm}
	\includegraphics[width=1.2\columnwidth]{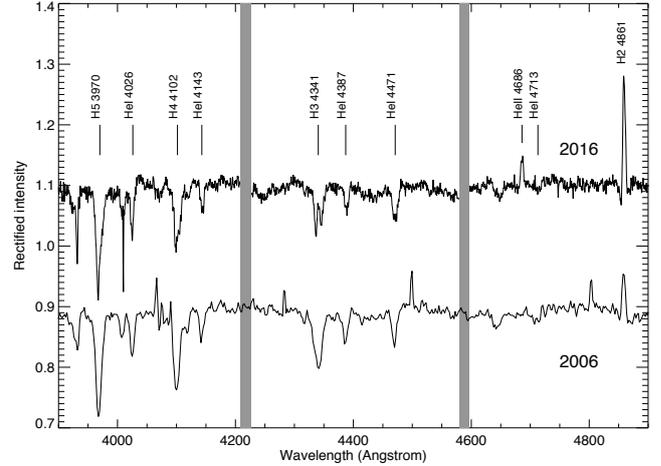}
    \caption{ESO 3.6\,m spectrum from 2006 and SALT RSS spectrum taken during the current X-ray outburst. The spectra are rectified and offset from 1.0 by 0.1 on either side for clarity. The vertical grey lines represent the chip gaps in the SALT CCD.}
    \label{fig:saltblue}
\end{figure}



\begin{figure*}
	\vspace{-1.5cm}
	\includegraphics[width=2.2\columnwidth]{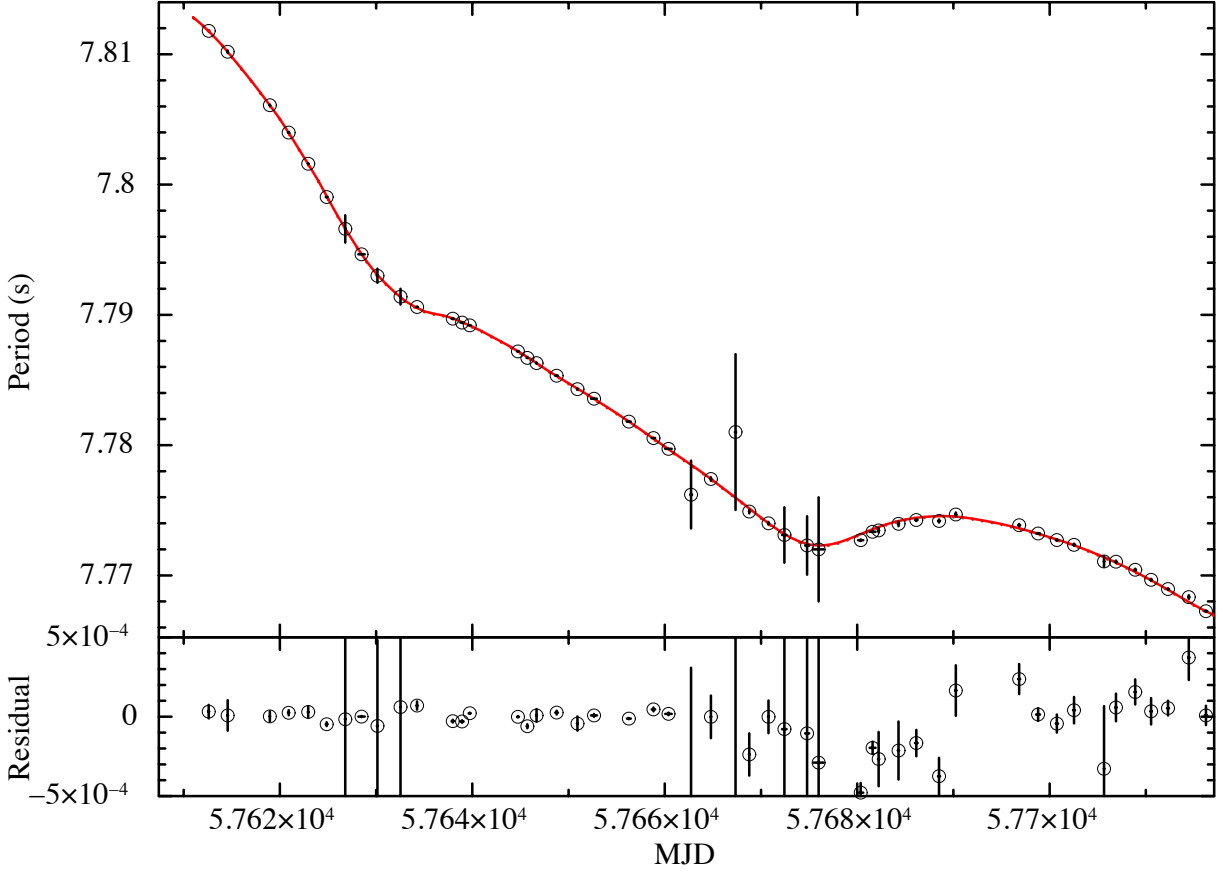}
	\vspace{-1.2cm}
    \caption{\textit{Top panel}: \textit{Swift} period measurements with the combined binary and spin-up model over-plotted in red. {Note that the fitted parameters for this model are given in Table~\ref{tab:orbitalfit} \textit{Lower panel}: Residuals of the fit.}}
    \label{fig:swiftper}
\end{figure*}

In addition to these data, we obtained a blue SALT/RSS spectrum to look for the presence of helium in emission. We compare this spectrum to an ESO 3.6\,m spectrum obtained in September 2006 (see \citealt{2008MNRAS.388.1198M} for a description of the ESO observation and the data reduction). The blue SALT/RSS spectrum was obtained using the PG2300 grating at an angle of 30.5 degrees, yielding a resolution of 2.1\,$\AA$. Fig.~\ref{fig:saltblue} shows the archival ESO spectrum and the recent SALT spectrum taken during the outburst. He\,II 4686 is clearly in emission during the outburst and not present in the archival spectrum (MJD 53993), which was taken during X-ray quiescence as shown by the lack of X-ray pulsations detected in \textit{RXTE} observations on MJD 53992 and 53998 (see also Fig.~\ref{fig:rxte}). This is likely evidence of a transient accretion disc around the neutron star. The H$_{\beta}$ line shows evidence of a very narrow emission component superimposed on a rotationally broadened absorption line, typical of many early-type emission line stars. The emission component is also stronger than it was in 2006. Emission in H$_{\gamma}$ and possibly H$_{\delta}$ is also apparent in the 2016 SALT spectrum. This is further good evidence that the disc has grown in the time between these observations.

\section{Orbital solution}

The measured pulsar period of SMC X-3 shows significant variability during the outburst, overall showing a decrease in period due to pulsar spin-up, but also showing modulation due to Doppler shift caused by orbital motion. We attempted to model the evolution of the pulsar period utilizing several models. Firstly we fit the data with a model comprising of a simple spin-up (i.e. constant $\dot{P}$), with orbital modulation. Unfortunately we were not able to obtain an adequate fit to the data utilizing this model, suggesting that the spin-up of the pulsar was not linear.

In an attempt to model this, we followed the example of \cite{2016PASJ...68S..13T} which modelled $\dot{P}$ utilizing the equation of \cite{1979ApJ...234..296G}, which suggests that the spin-up of a pulsar is proportional to L$_{37}^{6/7}$, where L$_{37}$ is the bolometric luminosity of the pulsar in units of 10$^{37}$ erg/s. We simplified this by assuming that all the values other than L$_{37}$ and P in equation 3 of \cite{2016PASJ...68S..13T} can be combined into a constant. Note that for L$_{37}$ we simply use a model fitted luminosity based on the spectral fitting described in section 2, and corrected for a standard SMC distance of 62\,kpc. We assume that the correction from this luminosity to a bolometric luminosity is constant, and include that in the fitted constant parameter. Therefore the fitted model becomes:

\begin{equation}
    \dot{P}=C \times P^{2} \times L_{37}(0.5-10$\,$\mathrm{keV})^{6/7}
	\label{eq:model}
\end{equation}
\smallskip

\noindent Fitting this model, including an orbital Doppler shift as detailed in \cite{2015MNRAS.447.2387C}, to our data provides a much improved fit over a simple linear $\dot{P}$, however the model underestimates the spin-up in the latter parts of the outburst. In order to compensate for this, we attempted three variations of the model. Firstly we added in an additional $\dot{P}$ value to account for the poor fit in the latter parts of the outburst. Secondly we allowed the luminosity index in equation \ref{eq:model} to be a fitted parameter. Thirdly we fit the model with a variable luminosity index and $\dot{P}$, which resulted in the best fitting solution as shown in Fig.~\ref{fig:swiftper} and Table~\ref{tab:orbitalfit}. We stress that this model is simply a parametrisation of the data, and does not represent a realistic model explaining the spin period evolution. However, in fitting the underlying period evolution like this, it allows us to accurately determine the orbital parameters of the system. It is apparent that the model struggles to fit the data where there are sharp or complex changes in the spin period (e.g. near MJD\,57675), which are likely due to some higher order spin variability that is difficult or impossible to model. Indeed, \cite{1979ApJ...234..296G} assume accretion is from an aligned disc, whereas in reality the accretion in a BeXRB is likely to be more complex than this, which may affect the relationship. The assumption that the correction from XRT flux to bolometric flux is a constant throughout the outburst is also likely not entirely accurate, given that the spectral parameters are seen to vary throughout the outburst.

In order to test whether our binary model does reliably determine the orbital parameters in this system, we applied a simple $\dot{P} + \ddot{P}$ model to the data that does not take into account the luminosity of the outburst. This removes the uncertainty arising from our assumption of a constant bolometric flux correction, but means that rapid changes in the spin period are less likely to be accounted for than in the previous model. Indeed, this proved to be the case as the model was unable to converge to a physically realistic set of parameters. We applied the same model to a subset of the data that excluded the initial super-Eddington part of the outburst, resulting in a much improved fit. This may be due to more complex spin period variability being present during the onset of the outburst, or because the errors on our spin period measurements are being underestimated at higher luminosities due to an imperfect model of the shape of the pulse profile. The result is marginally consistent with the solution in Table~\ref{tab:orbitalfit} (3\,$\sigma$), albeit with larger errors, and confirms that we are reliably extracting the true binary parameters in the full model.

\begin{table}
	\centering
	\caption{Binary solution obtained in this work.}
	\label{tab:orbitalfit}
	\begin{tabular}{|l|c|l|}
  		\hline
  		Parameter & & Value \\
  		\hline
  		\hline
  		Orbital period & $P_\mathrm{orbital}$ (d) & $45.04\pm0.08$ \\
  		Projected semimajor axis & $\textit{a}_{x}$sin{\it i} (light-s) & $190.3\pm1.3$ \\
  		Longitude of periastron & $\omega$ ($^{o}$) & $204.3\pm1.1$ \\
  		Eccentricity & $\textit{e}$ & $0.244\pm0.005$ \\
  		Orbital epoch & $\tau_\mathrm{per}$ (MJD) & $57676.4\pm0.2$ \\
  		GL79 index (c.f. 0.857) & GL79 index & $0.61\pm0.01$ \\
  		First derivative of $\textit{P}$ & $\dot{P}$ ($10^{-10}\mathrm{ss}^{-1}$) &  $-7.4\pm0.8$ \\
  		Goodness of fit & $\chi^{2}_{\nu}$ (d.o.f.) & 321 (54) \\
 		\hline
	\end{tabular}
\end{table}

\section{Discussion}

At the time of writing, SMC X-3 was still being detected at a high level of confidence with \textit{Swift} having been in X-ray outburst for 5 months. It is one of the longest and brightest outbursts ever recorded from a BeXRB. We discuss the results of our observations and compare SMC X-3 to the known population of BeXRBs.

\subsection{Neutron star spin} 

Fig.~\ref{fig:rxte} shows that the general spin-down trend observed in this system by \cite{2008ApJS..177..189G} continued through the latter stages of \textit{RXTE} monitoring, right up to the start of the current giant outburst on MJD 57599. At this point we detect significant spin-up, unlike anything observed in this system before. The spin period measured at the time of writing is lower than the very first measurement by \textit{RXTE}, showing that the momentum transferred by material accreted during the last 5 months has been greater than the momentum lost by magnetic breaking over the past 18 years. \cite{2014MNRAS.437.3863K} measure the long-term spin-down of SMC X-3 to be 0.00262 $\pm$ 0.00003\,s\,yr$^{-1}$ and the average luminosity (during outburst) to be (3.6 $\pm$ 0.1) $\times$ 10$^{36}$\,erg\,s$^{-1}$. This value is roughly 500 times lower than the spin-\textit{up} value of $\sim$\,$-$0.13\,s\,yr$^{-1}$ seen during the current outburst, showing the significantly larger torques present during this outburst. Even during previous Type I outbursts recorded by \textit{RXTE}, the spin period seems to continue increasing under low levels of accretion. This is probably evidence of extremely efficient accretion from a transient accretion disc in the current giant outburst.

\subsection{System parameters and mass function}

In trying to model the binary orbit of SMC X-3, we were using a dataset containing incredibly complex variations in the measured neutron star spin period. Thus, it was very difficult to properly account for the whole outburst without modifying certain accretion models or taking subsets of the data. The reason for not fixing the binary period at the very precise value determined from the optical light curve, is because of the growing evidence that this optical period does not necessarily represent the exact binary period (e.g. \citealt{2014A&A...567A.129V}, \citealt{2012MNRAS.423.3663B}). Whilst in this case we have shown that the periods are consistent at the 2\,$\sigma$ level, we cannot be certain enough that they are exactly the same to fix the parameter in the final fit. In addition, our binary parameters in Table~\ref{tab:orbitalfit} are entirely consistent with those found by \cite{2017arXiv170200966T} using a similar method, but show slight differences to those of \cite{2017arXiv170102983W} not consistent within errors.

The binary parameters are known in 7 other BeXRB systems in the SMC (\citealt{2011MNRAS.416.1556T}; \citealt{2015MNRAS.447.2387C}). When compared to this group, SMC X-3 seems to have a lower than expected eccentricity for its orbital period, sitting in between the 'normal' BeXRB systems and the group of low eccentricity systems suggested by \cite{2002ApJ...574..364P} in Fig.\,6 of \cite{2011MNRAS.416.1556T}. Thus, we may be starting to see evidence that these systems are formed from supernovae with a broad continuum of kick amplitudes, instead of the discrete 'large' or 'small' kick scenario, which results in a broad range of eccentricities and orbital periods. SMC X-3 also has the highest measured projected semi-major axis of the SMC systems to date, despite having a low eccentricity. The mass function derived from these parameters is 3.7\,M$_{\odot}$, typical of other BeXRBs.

\subsection{Circumstellar disc}

The stable and highly significant periodicity measured in the OGLE light curve is consistent with the binary period in Table~\ref{tab:orbitalfit} to within 2\,$\sigma$. This period is also seen in the long-term \textit{RXTE} light curve \citep{2008ApJS..177..189G}, making it probable that the optical period represents the binary period. We find an ephemeris of maximum optical flux of MJD\,57682.14 $\pm$ 0.37, whereas the dynamically determined orbital ephemeris is MJD\,57676.4 $\pm$ 0.2. This suggests that the peak in optical flux occurs roughly 6 days after periastron, and that this offset is stable over more than a decade. Smoothed-Particle Hydrodynamic (SPH) simulations \citep{2002MNRAS.337..967O} show that for any non-zero eccentricity the neutron star distorts the disc shape as it goes through periastron. This distortion increases the surface area of the circumstellar disc leading to an optical flare or enhancement. A lag of $\sim$\,6 days is an indication of how long this process takes, probably related to the disc viscosity. This effect is observed in other well-known systems, such as PSR\,B1259--63. The ephemeris derived from the \textit{RXTE} light curve is not precise enough to extrapolate forward and compare to the other ephemerides, so we do not know whether the maximum in X-ray flux occurs at periastron or not.

\begin{figure}
	\vspace{-1.4cm}
	\hspace{-0.6cm}
	\includegraphics[width=1.1\columnwidth,angle=0]{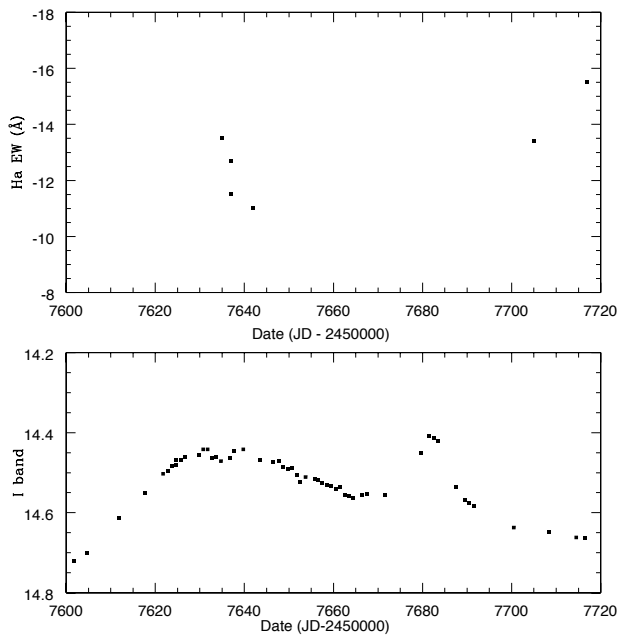}
	\vspace{-2.2cm}
    \caption{Measured H$_{\alpha}$ emission during the X-ray outburst of SMC X-3 plotting above the corresponding OGLE I-band photometry.}
    \label{fig:ha-ogle}
\end{figure}

The OGLE light curve and the historical H$_{\alpha}$ measurements suggest the disc in SMC X-3 has been very stable over more than a decade. Some of the Be stars in BeXRBs do exhibit this behaviour, though it is more common to see some large-scale variability on time-scales of a few years\footnote{See the OGLE XROM data analysis system \citep{2008AcA....58..187U} for long-term light curves of most BeXRBs in the SMC.}. The increase in optical flux coincides with the X-ray outburst, meaning the circumstellar disc is the source of extra material for this outburst, though the brightening of 0.3 mag is still only a moderate rise given the enormity of the X-ray outburst. The consistently single peaked H$_{\alpha}$ emission seems to point towards a disc that is not highly inclined to our line-of-sight, meaning a large increase in the size of the disc is not being hidden from us by a projected inclination effect.

Spectroscopic coverage of SMC X-3 has been sparse, making it hard to say whether the disc emission in H$_{\alpha}$ mimics the long-term broadband optical emission. However, we have observed the H$_{\alpha}$ emission to be variable on shorter time-scales, not seen in the optical light curve. Fig. \ref{fig:ha-ogle} shows the measured H$_{\alpha}$ emission during the outburst and the corresponding I-band flux during this time. One can see that the H$_{\alpha}$ emission changes by at least 1\,$\AA$ in a matter of hours and by as much as 50\,$\%$ in a few days. This variability is not reflected in the I-band flux. To check that the errors are not being underestimated, we measured the equivalent widths in several different ways: Gaussian fitting, randomised continuum fitting using a Monte-Carlo method and using the STARLINK/DIPSO software, where the errors are calculated using the prescriptions given by \cite{1986MNRAS.222..809H}. In all cases, the error bars were similar and confirmed the variability must be intrinsic to the source. Unfortunately, the coverage around any given orbit of the neutron star is not sufficient to say if this is linked to orbital phase in some way. The cause could be the interaction of the neutron star with inhomogeneities in the expanded disc, but this is difficult to confirm given these data.

\section{Conclusions}

This paper presents X-ray and optical observations made between 2016 July 30 and 2016 December 18 during the first 5 months of the giant X-ray outburst of SMC X-3. The peak X-ray luminosity is far in excess of the Eddington limit for a canonical mass neutron star, making it a nearby ultra-luminous X-ray source. The complex period changes, likely caused by variable accretion torques, have been untangled from the orbital modulations allowing us to measure the binary parameters. We show the binary parameters of this system are typical of other BeXRB systems, though perhaps somewhat on the edge of the distribution of eccentricities of 'normal' BeXRB systems given the longer binary period. When compared to historical spin period measurements made by \textit{RXTE}, the pulsar is seen to deviate from a state of constant spin-down to an extremely rapid spin-up that returned the neutron star spin period to that observed 18 years ago, in just 5 months.

The optical period of 44.918\,days is consistent with the dynamically determined binary period and the long-term X-ray period within errors. However, the ephemeris shows the optical emission is delayed by around 6 days from periastron, which may be linked to viscous time-scales in the disc as it is being distorted by the neutron star. The H$_{\alpha}$ emission from the disc is variable on short time-scales which may be linked to the disc being inhomogeneous or time-variable, though the data are not sufficient to confirm this hypothesis. We also observe He\,II in emission, showing the reservoir of material is sufficient to fuel a stable accretion disc around the neutron star, reflecting the enormity of this outburst.

\section*{Acknowledgements}

We would like to acknowledge Elizabeth Bartlett for providing the NTT spectrum included in our analysis. LJT is supported by the University of Cape Town Research Committee. JAK acknowledges the support of NASA grant NNX15AR44G through the Swift GI program. VAM and DAHB acknowledge support from the South African National Research Foundation. Some of these observations were obtained with the Southern African Large Telescope under program 2016-2-MLT-010. The OGLE project has received funding from the National Science Centre, Poland, grant MAESTRO 2014/14/A/ST9/00121 to AU.




\bibliographystyle{mnras}
\bibliography{smcx3} 

\begin{thebibliography}{}
\makeatletter
\relax
\def\mn@urlcharsother{\let\do\@makeother \do\$\do\&\do\#\do\^\do\_\do\%\do\~}
\def\mn@doi{\begingroup\mn@urlcharsother \@ifnextchar [ {\mn@doi@}
  {\mn@doi@[]}}
\def\mn@doi@[#1]#2{\def\@tempa{#1}\ifx\@tempa\@empty \href
  {http://dx.doi.org/#2} {doi:#2}\else \href {http://dx.doi.org/#2} {#1}\fi
  \endgroup}
\def\mn@eprint#1#2{\mn@eprint@#1:#2::\@nil}
\def\mn@eprint@arXiv#1{\href {http://arxiv.org/abs/#1} {{\tt arXiv:#1}}}
\def\mn@eprint@dblp#1{\href {http://dblp.uni-trier.de/rec/bibtex/#1.xml}
  {dblp:#1}}
\def\mn@eprint@#1:#2:#3:#4\@nil{\def\@tempa {#1}\def\@tempb {#2}\def\@tempc
  {#3}\ifx \@tempc \@empty \let \@tempc \@tempb \let \@tempb \@tempa \fi \ifx
  \@tempb \@empty \def\@tempb {arXiv}\fi \@ifundefined
  {mn@eprint@\@tempb}{\@tempb:\@tempc}{\expandafter \expandafter \csname
  mn@eprint@\@tempb\endcsname \expandafter{\@tempc}}}

\bibitem[\protect\citeauthoryear{{Antoniou}, {Zezas}, {Hatzidimitriou}  \&
  {Kalogera}}{{Antoniou} et~al.}{2010}]{2010ApJ...716L.140A}
{Antoniou} V.,  {Zezas} A.,  {Hatzidimitriou} D.,   {Kalogera} V.,  2010,
  \mn@doi [\apjl] {10.1088/2041-8205/716/2/L140}, \href
  {http://adsabs.harvard.edu/abs/2010ApJ...716L.140A} {716, L140}

\bibitem[\protect\citeauthoryear{{Bird}, {Coe}, {McBride}  \& {Udalski}}{{Bird}
  et~al.}{2012}]{2012MNRAS.423.3663B}
{Bird} A.~J.,  {Coe} M.~J.,  {McBride} V.~A.,   {Udalski} A.,  2012, \mn@doi
  [\mnras] {10.1111/j.1365-2966.2012.21163.x}, \href
  {http://adsabs.harvard.edu/abs/2012MNRAS.423.3663B} {423, 3663}

\bibitem[\protect\citeauthoryear{{Buccheri} et~al.,}{{Buccheri}
  et~al.}{1983}]{1983A&A...128..245B}
{Buccheri} R.,  et~al., 1983, \aap, \href
  {http://adsabs.harvard.edu/abs/1983A%26A...128..245B} {128, 245}

\bibitem[\protect\citeauthoryear{{Casares}, {Negueruela}, {Rib{\'o}}, {Ribas},
  {Paredes}, {Herrero}  \& {Sim{\'o}n-D{\'{\i}}az}}{{Casares}
  et~al.}{2014}]{2014Natur.505..378C}
{Casares} J.,  {Negueruela} I.,  {Rib{\'o}} M.,  {Ribas} I.,  {Paredes} J.~M.,
  {Herrero} A.,   {Sim{\'o}n-D{\'{\i}}az} S.,  2014, \mn@doi [\nat]
  {10.1038/nature12916}, \href
  {http://adsabs.harvard.edu/abs/2014Natur.505..378C} {505, 378}

\bibitem[\protect\citeauthoryear{{Clark}, {Doxsey}, {Li}, {Jernigan}  \& {van
  Paradijs}}{{Clark} et~al.}{1978}]{1978ApJ...221L..37C}
{Clark} G.,  {Doxsey} R.,  {Li} F.,  {Jernigan} J.~G.,   {van Paradijs} J.,
  1978, \mn@doi [\apjl] {10.1086/182660}, \href
  {http://cdsads.u-strasbg.fr/abs/1978ApJ...221L..37C} {221, L37}

\bibitem[\protect\citeauthoryear{{Coe} \& {Kirk}}{{Coe} \&
  {Kirk}}{2015}]{2015MNRAS.452..969C}
{Coe} M.~J.,  {Kirk} J.,  2015, \mn@doi [\mnras] {10.1093/mnras/stv1283}, \href
  {http://adsabs.harvard.edu/abs/2015MNRAS.452..969C} {452, 969}

\bibitem[\protect\citeauthoryear{{Coe}, {Bartlett}, {Bird}, {Haberl}, {Kennea},
  {McBride}, {Townsend}  \& {Udalski}}{{Coe}
  et~al.}{2015}]{2015MNRAS.447.2387C}
{Coe} M.~J.,  {Bartlett} E.~S.,  {Bird} A.~J.,  {Haberl} F.,  {Kennea} J.~A.,
  {McBride} V.~A.,  {Townsend} L.~J.,   {Udalski} A.,  2015, \mn@doi [\mnras]
  {10.1093/mnras/stu2568}, \href
  {http://adsabs.harvard.edu/abs/2015MNRAS.447.2387C} {447, 2387}

\bibitem[\protect\citeauthoryear{{Corbet}, {Edge}, {Laycock}, {Coe},
  {Markwardt}  \& {Marshall}}{{Corbet} et~al.}{2003}]{2003HEAD....7.1730C}
{Corbet} R.~H.~D.,  {Edge} W.~R.~T.,  {Laycock} S.,  {Coe} M.~J.,  {Markwardt}
  C.~B.,   {Marshall} F.~E.,  2003, in AAS/High Energy Astrophysics Division
  \#7. p.~629

\bibitem[\protect\citeauthoryear{{Cowley} \& {Schmidtke}}{{Cowley} \&
  {Schmidtke}}{2004}]{2004AJ....128..709C}
{Cowley} A.~P.,  {Schmidtke} P.~C.,  2004, \mn@doi [\aj] {10.1086/422025},
  \href {http://cdsads.u-strasbg.fr/abs/2004AJ....128..709C} {128, 709}

\bibitem[\protect\citeauthoryear{{Crampton}, {Hutchings}  \&
  {Cowley}}{{Crampton} et~al.}{1978}]{1978ApJ...223L..79C}
{Crampton} D.,  {Hutchings} J.~B.,   {Cowley} A.~P.,  1978, \mn@doi [\apjl]
  {10.1086/182733}, \href {http://cdsads.u-strasbg.fr/abs/1978ApJ...223L..79C}
  {223, L79}

\bibitem[\protect\citeauthoryear{{Cusumano} et~al.,}{{Cusumano}
  et~al.}{2012}]{2012A&A...548A..28C}
{Cusumano} G.,  et~al., 2012, \mn@doi [\aap] {10.1051/0004-6361/201219968},
  \href {http://adsabs.harvard.edu/abs/2012A%26A...548A..28C} {548, A28}

\bibitem[\protect\citeauthoryear{{Dray}}{{Dray}}{2006}]{2006MNRAS.370.2079D}
{Dray} L.~M.,  2006, \mn@doi [\mnras] {10.1111/j.1365-2966.2006.10635.x}, \href
  {http://adsabs.harvard.edu/abs/2006MNRAS.370.2079D} {370, 2079}

\bibitem[\protect\citeauthoryear{{Edge}}{{Edge}}{2005}]{2005PhDT.........3E}
{Edge} W.~R.~T.,  2005, PhD thesis, University of Southampton (United Kingdom),
  England

\bibitem[\protect\citeauthoryear{{Edge}, {Coe}, {Corbet}, {Markwardt}  \&
  {Laycock}}{{Edge} et~al.}{2004}]{2004ATel..225....1E}
{Edge} W.~R.~T.,  {Coe} M.~J.,  {Corbet} R.~H.~D.,  {Markwardt} C.~B.,
  {Laycock} S.,  2004, The Astronomer's Telegram, \href
  {http://cdsads.u-strasbg.fr/abs/2004ATel..225....1E} {225}

\bibitem[\protect\citeauthoryear{{Galache}, {Corbet}, {Coe}, {Laycock},
  {Schurch}, {Markwardt}, {Marshall}  \& {Lochner}}{{Galache}
  et~al.}{2008}]{2008ApJS..177..189G}
{Galache} J.~L.,  {Corbet} R.~H.~D.,  {Coe} M.~J.,  {Laycock} S.,  {Schurch}
  M.~P.~E.,  {Markwardt} C.,  {Marshall} F.~E.,   {Lochner} J.,  2008, \mn@doi
  [\apjs] {10.1086/587743}, \href
  {http://cdsads.u-strasbg.fr/abs/2008ApJS..177..189G} {177, 189}

\bibitem[\protect\citeauthoryear{{Ghosh} \& {Lamb}}{{Ghosh} \&
  {Lamb}}{1979}]{1979ApJ...234..296G}
{Ghosh} P.,  {Lamb} F.~K.,  1979, \mn@doi [\apj] {10.1086/157498}, \href
  {http://adsabs.harvard.edu/abs/1979ApJ...234..296G} {234, 296}

\bibitem[\protect\citeauthoryear{{Gotthelf}, {Vasisht}  \& {Dotani}}{{Gotthelf}
  et~al.}{1999}]{1999ApJ...522L..49G}
{Gotthelf} E.~V.,  {Vasisht} G.,   {Dotani} T.,  1999, \mn@doi [\apjl]
  {10.1086/312220}, \href {http://adsabs.harvard.edu/abs/1999ApJ...522L..49G}
  {522, L49}

\bibitem[\protect\citeauthoryear{{Haberl} \& {Sturm}}{{Haberl} \&
  {Sturm}}{2016}]{2016A&A...586A..81H}
{Haberl} F.,  {Sturm} R.,  2016, \mn@doi [\aap] {10.1051/0004-6361/201527326},
  \href {http://adsabs.harvard.edu/abs/2016A%26A...586A..81H} {586, A81}

\bibitem[\protect\citeauthoryear{{Haberl}, {Eger}  \& {Pietsch}}{{Haberl}
  et~al.}{2008}]{2008A&A...489..327H}
{Haberl} F.,  {Eger} P.,   {Pietsch} W.,  2008, \mn@doi [\aap]
  {10.1051/0004-6361:200810100}, \href
  {http://cdsads.u-strasbg.fr/abs/2008A%26A...489..327H} {489, 327}

\bibitem[\protect\citeauthoryear{{Horne} \& {Baliunas}}{{Horne} \&
  {Baliunas}}{1986}]{1986ApJ...302..757H}
{Horne} J.~H.,  {Baliunas} S.~L.,  1986, \mn@doi [\apj] {10.1086/164037}, \href
  {http://adsabs.harvard.edu/abs/1986ApJ...302..757H} {302, 757}

\bibitem[\protect\citeauthoryear{{Howarth} \& {Phillips}}{{Howarth} \&
  {Phillips}}{1986}]{1986MNRAS.222..809H}
{Howarth} I.~D.,  {Phillips} A.~P.,  1986, \mn@doi [\mnras]
  {10.1093/mnras/222.4.809}, \href
  {http://adsabs.harvard.edu/abs/1986MNRAS.222..809H} {222, 809}

\bibitem[\protect\citeauthoryear{{Kennea} et~al.,}{{Kennea}
  et~al.}{2016}]{2016ATel.9362....1K}
{Kennea} J.~A.,  et~al., 2016, The Astronomer's Telegram, \href
  {http://adsabs.harvard.edu/abs/2016ATel.9362....1K} {9362}

\bibitem[\protect\citeauthoryear{{Klus}, {Ho}, {Coe}, {Corbet}  \&
  {Townsend}}{{Klus} et~al.}{2014}]{2014MNRAS.437.3863K}
{Klus} H.,  {Ho} W.~C.~G.,  {Coe} M.~J.,  {Corbet} R.~H.~D.,   {Townsend}
  L.~J.,  2014, \mn@doi [\mnras] {10.1093/mnras/stt2192}, \href
  {http://adsabs.harvard.edu/abs/2014MNRAS.437.3863K} {437, 3863}

\bibitem[\protect\citeauthoryear{{Krimm} et~al.,}{{Krimm}
  et~al.}{2013}]{2013ApJS..209...14K}
{Krimm} H.~A.,  et~al., 2013, \mn@doi [\apjs] {10.1088/0067-0049/209/1/14},
  \href {http://adsabs.harvard.edu/abs/2013ApJS..209...14K} {209, 14}

\bibitem[\protect\citeauthoryear{{Li}, {Jernigan}  \& {Clark}}{{Li}
  et~al.}{1977}]{1977IAUC.3125....1L}
{Li} F.,  {Jernigan} G.,   {Clark} G.,  1977, \iaucirc, \href
  {http://cdsads.u-strasbg.fr/abs/1977IAUC.3125....1L} {3125}

\bibitem[\protect\citeauthoryear{{McBride}, {Coe}, {Negueruela}, {Schurch}  \&
  {McGowan}}{{McBride} et~al.}{2008}]{2008MNRAS.388.1198M}
{McBride} V.~A.,  {Coe} M.~J.,  {Negueruela} I.,  {Schurch} M.~P.~E.,
  {McGowan} K.~E.,  2008, \mn@doi [\mnras] {10.1111/j.1365-2966.2008.13410.x},
  \href {http://cdsads.u-strasbg.fr/abs/2008MNRAS.388.1198M} {388, 1198}

\bibitem[\protect\citeauthoryear{{Negoro} et~al.,}{{Negoro}
  et~al.}{2016}]{2016ATel.9348....1N}
{Negoro} H.,  et~al., 2016, The Astronomer's Telegram, \href
  {http://adsabs.harvard.edu/abs/2016ATel.9348....1N} {9348}

\bibitem[\protect\citeauthoryear{{Okazaki}, {Bate}, {Ogilvie}  \&
  {Pringle}}{{Okazaki} et~al.}{2002}]{2002MNRAS.337..967O}
{Okazaki} A.~T.,  {Bate} M.~R.,  {Ogilvie} G.~I.,   {Pringle} J.~E.,  2002,
  \mn@doi [\mnras] {10.1046/j.1365-8711.2002.05960.x}, \href
  {http://adsabs.harvard.edu/abs/2002MNRAS.337..967O} {337, 967}

\bibitem[\protect\citeauthoryear{{Pfahl}, {Rappaport}, {Podsiadlowski}  \&
  {Spruit}}{{Pfahl} et~al.}{2002}]{2002ApJ...574..364P}
{Pfahl} E.,  {Rappaport} S.,  {Podsiadlowski} P.,   {Spruit} H.,  2002, \mn@doi
  [\apj] {10.1086/340794}, \href
  {http://adsabs.harvard.edu/abs/2002ApJ...574..364P} {574, 364}

\bibitem[\protect\citeauthoryear{{Reig}}{{Reig}}{2011}]{2011Ap&SS.332....1R}
{Reig} P.,  2011, \mn@doi [\apss] {10.1007/s10509-010-0575-8}, \href
  {http://adsabs.harvard.edu/abs/2011Ap%26SS.332....1R} {332, 1}

\bibitem[\protect\citeauthoryear{{Schmidtke}, {Cowley}  \&
  {Udalski}}{{Schmidtke} et~al.}{2013}]{2013MNRAS.431..252S}
{Schmidtke} P.~C.,  {Cowley} A.~P.,   {Udalski} A.,  2013, \mn@doi [\mnras]
  {10.1093/mnras/stt159}, \href
  {http://cdsads.u-strasbg.fr/abs/2013MNRAS.431..252S} {431, 252}

\bibitem[\protect\citeauthoryear{{Scowcroft}, {Freedman}, {Madore}, {Monson},
  {Persson}, {Rich}, {Seibert}  \& {Rigby}}{{Scowcroft}
  et~al.}{2016}]{2016ApJ...816...49S}
{Scowcroft} V.,  {Freedman} W.~L.,  {Madore} B.~F.,  {Monson} A.,  {Persson}
  S.~E.,  {Rich} J.,  {Seibert} M.,   {Rigby} J.~R.,  2016, \mn@doi [\apj]
  {10.3847/0004-637X/816/2/49}, \href
  {http://adsabs.harvard.edu/abs/2016ApJ...816...49S} {816, 49}

\bibitem[\protect\citeauthoryear{{Takagi}, {Mihara}, {Sugizaki}, {Makishima}
  \& {Morii}}{{Takagi} et~al.}{2016}]{2016PASJ...68S..13T}
{Takagi} T.,  {Mihara} T.,  {Sugizaki} M.,  {Makishima} K.,   {Morii} M.,
  2016, \mn@doi [\pasj] {10.1093/pasj/psw010}, \href
  {http://adsabs.harvard.edu/abs/2016PASJ...68S..13T} {68, S13}

\bibitem[\protect\citeauthoryear{{Townsend}, {Coe}, {Corbet}  \&
  {Hill}}{{Townsend} et~al.}{2011}]{2011MNRAS.416.1556T}
{Townsend} L.~J.,  {Coe} M.~J.,  {Corbet} R.~H.~D.,   {Hill} A.~B.,  2011,
  \mn@doi [\mnras] {10.1111/j.1365-2966.2011.19153.x}, \href
  {http://adsabs.harvard.edu/abs/2011MNRAS.416.1556T} {416, 1556}

\bibitem[\protect\citeauthoryear{{Tsygankov}, {Doroshenko}, {Lutovinov},
  {Mushtukov}  \& {Poutanen}}{{Tsygankov} et~al.}{2017}]{2017arXiv170200966T}
{Tsygankov} S.~S.,  {Doroshenko} V.,  {Lutovinov} A.~A.,  {Mushtukov} A.~A.,
  {Poutanen} J.,  2017, preprint, \href
  {http://adsabs.harvard.edu/abs/2017arXiv170200966T} {} (\mn@eprint {arXiv}
  {1702.00966})

\bibitem[\protect\citeauthoryear{{Udalski}}{{Udalski}}{2008}]{2008AcA....58..187U}
{Udalski} A.,  2008, \actaa, \href
  {http://adsabs.harvard.edu/abs/2008AcA....58..187U} {58, 187}

\bibitem[\protect\citeauthoryear{{Udalski}, {Szymanski}, {Soszynski}  \&
  {Poleski}}{{Udalski} et~al.}{2008}]{2008AcA....58...69U}
{Udalski} A.,  {Szymanski} M.~K.,  {Soszynski} I.,   {Poleski} R.,  2008,
  \actaa, \href {http://adsabs.harvard.edu/abs/2008AcA....58...69U} {58, 69}

\bibitem[\protect\citeauthoryear{{Udalski}, {Szyma{\'n}ski}  \&
  {Szyma{\'n}ski}}{{Udalski} et~al.}{2015}]{2015AcA....65....1U}
{Udalski} A.,  {Szyma{\'n}ski} M.~K.,   {Szyma{\'n}ski} G.,  2015, \actaa,
  \href {http://adsabs.harvard.edu/abs/2015AcA....65....1U} {65, 1}

\bibitem[\protect\citeauthoryear{{Vasilopoulos}, {Haberl}, {Sturm}, {Maggi}  \&
  {Udalski}}{{Vasilopoulos} et~al.}{2014}]{2014A&A...567A.129V}
{Vasilopoulos} G.,  {Haberl} F.,  {Sturm} R.,  {Maggi} P.,   {Udalski} A.,
  2014, \mn@doi [\aap] {10.1051/0004-6361/201423934}, \href
  {http://adsabs.harvard.edu/abs/2014A%26A...567A.129V} {567, A129}

\bibitem[\protect\citeauthoryear{{Vasilopoulos}, {Haberl}, {Antoniou}  \&
  {Zezas}}{{Vasilopoulos} et~al.}{2016}]{2016ATel.9229....1V}
{Vasilopoulos} G.,  {Haberl} F.,  {Antoniou} V.,   {Zezas} A.,  2016, The
  Astronomer's Telegram, \href
  {http://adsabs.harvard.edu/abs/2016ATel.9229....1V} {9229}

\bibitem[\protect\citeauthoryear{{Walter}, {Lutovinov}, {Bozzo}  \&
  {Tsygankov}}{{Walter} et~al.}{2015}]{2015A&ARv..23....2W}
{Walter} R.,  {Lutovinov} A.~A.,  {Bozzo} E.,   {Tsygankov} S.~S.,  2015,
  \mn@doi [\aapr] {10.1007/s00159-015-0082-6}, \href
  {http://adsabs.harvard.edu/abs/2015A%26ARv..23....2W} {23, 2}

\bibitem[\protect\citeauthoryear{{Weng}, {Ge}, {Zhao}, {Wang}, {Zhang}, {Bian}
  \& {Yuan}}{{Weng} et~al.}{2017}]{2017arXiv170102983W}
{Weng} S.-S.,  {Ge} M.-Y.,  {Zhao} H.-H.,  {Wang} W.,  {Zhang} S.-N.,  {Bian}
  W.-H.,   {Yuan} Q.-R.,  2017, preprint, \href
  {http://adsabs.harvard.edu/abs/2017arXiv170102983W} {} (\mn@eprint {arXiv}
  {1701.02983})

\bibitem[\protect\citeauthoryear{{van Paradijs}, {Schlosser}, {Tarenghi},
  {Sanduleak}  \& {Philip}}{{van Paradijs} et~al.}{1977}]{1977IAUC.3134....3V}
{van Paradijs} J.,  {Schlosser} W.,  {Tarenghi} M.,  {Sanduleak} N.,   {Philip}
  A.~G.~D.,  1977, \iaucirc, \href
  {http://adsabs.harvard.edu/abs/1977IAUC.3134....3V} {3134}

\makeatother
\end{thebibliography}




\appendix

\section{Details of \textit{Swift} observations}


Table \ref{tab:swiftapp} presents the OBSID, exposure time, measured luminosity and measured spin period for each \textit{Swift} observation used in this work.
\begin{table}
\begin{center}
\caption{\textit{Swift} observations used in this work.}
	\label{tab:swiftapp}
\resizebox{1.2\columnwidth}{!}{%
\begin{tabular}{cccccc}
\hline
ObsID &T$_\mathrm{START}$ &$T_\mathrm{STOP}$ & Exposure & 0.5--10 keV Luminosity & Period \\
      &(MJD)      &(MJD)    & (s)      & $(10^{38}~\mathrm{erg/s})$ & (s) \\
\hline
00034673001 &57610.93 &57611.07 & 4658 & $3.85^{+0.05}_{-0.05}$ & $--$ \\
00034673001 &57611.00 &57611.00 & 75 & $0.33^{+0.26}_{-0.17}$ & $--$ \\
00034673002 &57612.53 &57612.66 & 1971 & $4.72^{+0.08}_{-0.08}$ & $7.811800 \pm 0.000043$ \\
00034673003 &57614.53 &57614.60 & 932 & $6.38^{+0.14}_{-0.14}$ & $7.810200 \pm 0.000096$ \\
00034673004 &57618.91 &57618.98 & 1993 & $6.69^{+0.08}_{-0.08}$ & $7.806100 \pm 0.000032$ \\
00034673005 &57620.84 &57620.98 & 2044 & $8.77^{+0.12}_{-0.12}$ & $7.804000 \pm 0.000016$ \\
00034673006 &57622.90 &57622.97 & 1953 & $9.00^{+0.10}_{-0.10}$ & $7.801600 \pm 0.000036$ \\
00034673007 &57624.76 &57624.96 & 1287 & $11.44^{+0.15}_{-0.15}$ & $7.799046 \pm 0.000012$ \\
00034673008 &57626.75 &57626.82 & 503 & $10.30^{+0.22}_{-0.22}$ & $7.796600 \pm 0.001049$ \\
00034673009 &57628.15 &57628.81 & 1578 & $10.43^{+0.13}_{-0.13}$ & $7.794653 \pm 0.000004$ \\
00034673010 &57630.07 &57630.19 & 795 & $9.76^{+0.18}_{-0.17}$ & $7.793000 \pm 0.000540$ \\
00034673011 &57632.52 &57632.53 & 792 & $10.81^{+0.21}_{-0.21}$ & $7.791400 \pm 0.000605$ \\
00034673012 &57634.19 &57634.32 & 1925 & $8.46^{+0.12}_{-0.11}$ & $7.790600 \pm 0.000026$ \\
00034673013 &57636.17 &57636.64 & 1339 & $5.15^{+0.10}_{-0.10}$ & $7.788500 \pm 0.000777$ \\
00034673014 &57637.83 &57638.09 & 2974 & $5.97^{+0.06}_{-0.06}$ & $7.789706 \pm 0.000009$ \\
00034673015 &57638.76 &57639.09 & 2963 & $6.62^{+0.08}_{-0.08}$ & $7.789408 \pm 0.000008$ \\
00034673016 &57639.62 &57639.83 & 2756 & $5.63^{+0.07}_{-0.07}$ & $7.789200 \pm 0.000004$ \\
00034673017 &57644.64 &57644.85 & 2966 & $4.83^{+0.06}_{-0.06}$ & $7.787200 \pm 0.000004$ \\
00034673018 &57645.63 &57645.78 & 2973 & $4.54^{+0.06}_{-0.06}$ & $7.786700 \pm 0.000013$ \\
00034673019 &57646.63 &57646.71 & 2975 & $4.67^{+0.06}_{-0.06}$ & $7.786300 \pm 0.000040$ \\
00034673020 &57648.62 &57648.90 & 2995 & $3.47^{+0.05}_{-0.05}$ & $7.785329 \pm 0.000012$ \\
00034673021 &57650.87 &57650.96 & 2912 & $3.02^{+0.05}_{-0.05}$ & $7.784300 \pm 0.000045$ \\
00034673022 &57652.33 &57652.95 & 2583 & $2.91^{+0.05}_{-0.05}$ & $7.783560 \pm 0.000009$ \\
00034673023 &57654.01 &57654.68 & 2982 & $2.70^{+0.04}_{-0.04}$ & $7.783964 \pm 0.000010$ \\
00034673024 &57656.06 &57656.46 & 2070 & $2.60^{+0.05}_{-0.05}$ & $7.781800 \pm 0.000004$ \\
00034673025 &57658.64 &57659.00 & 3279 & $2.42^{+0.04}_{-0.04}$ & $7.780547 \pm 0.000012$ \\
00034673026 &57660.05 &57660.72 & 2959 & $2.03^{+0.04}_{-0.04}$ & $7.779710 \pm 0.000008$ \\
00034673027 &57662.63 &57662.83 & 550 & $1.61^{+0.09}_{-0.09}$ & $7.776200 \pm 0.002603$ \\
00034673028 &57664.75 &57664.83 & 1713 & $1.62^{+0.05}_{-0.05}$ & $7.777400 \pm 0.000134$ \\
00034673029 &57667.28 &57667.42 & 723 & $1.38^{+0.10}_{-0.09}$ & $7.781000 \pm 0.005977$ \\
00034673030 &57668.75 &57668.82 & 2783 & $1.53^{+0.04}_{-0.04}$ & $7.774900 \pm 0.000133$ \\
00034673031 &57670.73 &57670.81 & 2833 & $1.33^{+0.03}_{-0.03}$ & $7.774000 \pm 0.000103$ \\
00034673032 &57672.19 &57672.66 & 1976 & $1.27^{+0.04}_{-0.04}$ & $7.773100 \pm 0.002129$ \\
00034673033 &57674.58 &57675.00 & 2020 & $1.21^{+0.04}_{-0.04}$ & $7.772300 \pm 0.002243$ \\
00034673034 &57675.46 &57676.52 & 1195 & $1.10^{+0.05}_{-0.05}$ & $7.772000 \pm 0.003997$ \\
00034673035 &57680.10 &57680.63 & 2483 & $1.00^{+0.04}_{-0.03}$ & $7.772697 \pm 0.000061$ \\
00034673036 &57681.30 &57681.90 & 2844 & $0.95^{+0.03}_{-0.03}$ & $7.773350 \pm 0.000039$ \\
00034673037 &57682.16 &57682.29 & 3426 & $0.94^{+0.03}_{-0.03}$ & $7.773456 \pm 0.000171$ \\
00034673038 &57684.21 &57684.36 & 3677 & $0.94^{+0.03}_{-0.03}$ & $7.773963 \pm 0.000183$ \\
00034673039 &57686.01 &57686.28 & 3615 & $0.86^{+0.03}_{-0.03}$ & $7.774258 \pm 0.000083$ \\
00034673040 &57688.40 &57688.61 & 2939 & $0.74^{+0.03}_{-0.03}$ & $7.774179 \pm 0.000116$ \\
00034673041 &57690.18 &57690.33 & 3976 & $0.91^{+0.03}_{-0.03}$ & $7.774677 \pm 0.000159$ \\
00034673043 &57696.73 &57696.94 & 3963 & $0.77^{+0.03}_{-0.03}$ & $7.773852 \pm 0.000095$ \\
00034673042 &57698.59 &57699.00 & 5241 & $0.74^{+0.02}_{-0.02}$ & $7.773220 \pm 0.000040$ \\
00034673044 &57700.65 &57700.86 & 4636 & $0.60^{+0.02}_{-0.02}$ & $7.772715 \pm 0.000057$ \\
00034673045 &57702.44 &57702.64 & 4441 & $0.63^{+0.02}_{-0.02}$ & $7.772353 \pm 0.000083$ \\
00034673046 &57706.82 &57706.97 & 4439 & $0.58^{+0.02}_{-0.02}$ & $7.771050 \pm 0.000087$ \\
00034673047 &57708.81 &57708.97 & 4318 & $0.57^{+0.02}_{-0.02}$ & $7.770433 \pm 0.000079$ \\
00034673048 &57710.47 &57710.62 & 4617 & $0.53^{+0.02}_{-0.02}$ & $7.769663 \pm 0.000083$ \\
00034673049 &57712.20 &57712.41 & 4869 & $0.51^{+0.02}_{-0.02}$ & $7.768950 \pm 0.000047$ \\
00034673050 &57714.40 &57714.54 & 3278 & $0.45^{+0.03}_{-0.02}$ & $7.768338 \pm 0.000142$ \\
00034673051 &57716.11 &57716.39 & 4472 & $0.39^{+0.02}_{-0.02}$ & $7.767253 \pm 0.000061$ \\
00034673053 &57718.31 &57718.46 & 4421 & $0.37^{+0.02}_{-0.02}$ & $7.766753 \pm 0.000096$ \\
00034673054 &57720.23 &57720.45 & 4578 & $0.30^{+0.02}_{-0.02}$ & $7.766453 \pm 0.000082$ \\
00034673055 &57722.55 &57722.77 & 5014 & $0.33^{+0.02}_{-0.02}$ & $7.766975 \pm 0.000084$ \\
00034673056 &57724.08 &57724.30 & 4552 & $0.26^{+0.02}_{-0.02}$ & $7.767576 \pm 0.000060$ \\
00034673057 &57726.54 &57726.76 & 4903 & $0.23^{+0.02}_{-0.02}$ & $7.768563 \pm 0.000081$ \\
00034673058 &57728.53 &57728.75 & 4906 & $0.16^{+0.01}_{-0.01}$ & $7.769552 \pm 0.000079$ \\
00034673059 &57730.07 &57730.48 & 4513 & $0.22^{+0.02}_{-0.02}$ & $7.769852 \pm 0.000071$ \\
00034673060 &57732.05 &57732.20 & 4700 & $0.26^{+0.01}_{-0.01}$ & $7.770529 \pm 0.000095$ \\
00034673061 &57734.79 &57734.98 & 3064 & $0.26^{+0.02}_{-0.02}$ & $7.770563 \pm 0.000108$ \\
00034673062 &57736.44 &57736.72 & 3529 & $0.26^{+0.02}_{-0.02}$ & $7.770734 \pm 0.000071$ \\
00034673063 &57738.03 &57738.31 & 4367 & $0.29^{+0.02}_{-0.02}$ & $7.770563 \pm 0.000050$ \\
00034673064 &57750.42 &57750.96 & 5024 & $0.22^{+0.01}_{-0.01}$ & $7.768773 \pm 0.000027$ \\
00034673065 &57752.54 &57752.74 & 4279 & $0.20^{+0.01}_{-0.01}$ & $7.768268 \pm 0.000120$ \\
00034673066 &57754.20 &57754.41 & 4493 & $0.18^{+0.01}_{-0.01}$ & $7.767672 \pm 0.000116$ \\
00034673067 &57763.42 &57763.96 & 1484 & $0.08^{+0.02}_{-0.02}$ & $--$ \\
00034673069 &57766.21 &57766.54 & 1660 & $0.04^{+0.01}_{-0.01}$ & $--$ \\
00034673081 &57780.10 &57780.31 & 2289 & $0.17^{+0.02}_{-0.02}$ & $--$ \\
\hline
\end{tabular}%
}
\end{center}
\end{table}


\bsp	
\label{lastpage}
\end{document}